\documentclass[aps,prl,twocolumn]{revtex4}
\usepackage{graphicx}
\usepackage{color}
\usepackage{dcolumn}

\newcommand{\EG}{{\textrm{e.g.}}}
\newcommand{\IE}{{\textrm{i.e.}}}
\newcommand{\EA}{{\textit{et al.}}}

\date{\today}

\begin{document}
NT@UW-15-02
\title{Comment on ``Neutron Skin of $^{208}$Pb from Coherent Pion Photoproduction''}

\author{A. G{\aa}rdestig}\email{gardestigae@bethanylb.edu}
\affiliation{Department of Physics, Bethany College, Lindsborg, KS 67456, U.S.A.}
\author{C. J. Horowitz}\email{horowit@indiana.edu}
\affiliation{Department of Physics and Nuclear Theory Center, Indiana University, Bloomington, IN 47408, U.S.A.}
\author{Gerald A. Miller}\email{miller@phys.washington.edu}
\affiliation{Department of Physics,	University of Washington, Seattle, Washington 98195-1560, U.S.A.}

\maketitle{}

In a recent article~\cite{CBMAMI}, the Crystal Ball at MAMI and A2 collaboration claims to have measured the root-mean-square (rms) neutron-proton radius difference $\Delta r_{np}$ of $^{208}$Pb to 0.15 fm $\pm0.03$ fm (stat.)  $+0.01$ or $-0.03$ fm (syst.) using coherent $(\gamma,\pi^0)$ photoproduction. 
This extraction uses a state-of-the-art single-nucleon $(\gamma,\pi^0)$ amplitude \cite{Drechsel} and a pion rescattering wave function computed using an optical potential. 
However, it does not include effects contributing to pion photoproduction, such as pion charge exchange scattering and medium modifications of nucleon resonances. 
We perform a simple calculation to show that higher order corrections to the $q^2$ dependence of the $(\gamma,\pi^0)$ amplitude that they use have to be known to within about 3\%. 
Such a high accuracy seems to be well beyond the convergence of any chiral expansion or the accuracy of any present phenomenological model.

The neutral photopion production reaction is sensitive to the sum of neutron and proton densities. 
The proton density is well-determined  by electron scattering, so  the  analysis of  coherent $(\gamma,\pi^0)$ photoproduction process may focus on the neutron density.  
In the following we simplify the notation by simply referring to a single density, $\rho_A$.

If one neglects the effects of  pion rescattering, the matrix element for coherent pion photo-production on $^{208}$Pb is expressed as the product $M=F_{\gamma\pi^0}\rho_A(q^2)$, where $F_{\gamma\pi^0}$ is the pion photo-production amplitude and $\rho_A(q^2)$ the relevant nuclear form factor, i.e., the Fourier transform of the nuclear density at a squared momentum transfer of $q^2$. 
For small values of $q^2$, the nuclear form factor can be expanded as $\rho_B(q^2) = 1+\frac{q^2R_A^2}{6}$, where $R_A$ is the  nuclear rms radius. 
The effects of pion-nuclear interactions can be of thought as modifying the amplitude $F_{\gamma\pi^0}$ to an {\it effective} amplitude, so that 
$F^{eff}_{\gamma\pi^0} = F_{\gamma\pi^0}(q^2)\left(1+a'\frac{q^2}{m_\pi^2}\right)$,
where $a'$ parametrizes the  leading correction to the single-nucleon production operator.

Our concern here is with the uncertainties in the value of $a'$. These arise from at least four sources:
the uncertainty in the single-nucleon amplitude of Ref.~\cite{CBMAMI} adapted from Ref.~\cite{Drechsel} arising from the fitted data,
the uncertainty in the optical potential (related to lack of knowledge of the pion wave function in the nuclear interior \cite{Miller:1974nm,Keister:1978ey}),
the neglected charge-exchange effects that are known to be important for light targets~\cite{EW},
and pion production on multiple nucleons.  
We estimate that the charge exchange effects are important for heavy targets also. 
The relevant amplitude is one in which $\gamma +n+p\rightarrow \pi^- +p+p\rightarrow \pi^0 +p +n$. 
There is also a term in which the intermediate pion is $\pi^++n+n$.  We note that this process is coherent and depends on the number of neutron-proton pairs in the nucleus.

More generally, one must consider all contributions involving  pion production on two or more nucleons.  
It is not enough that the amplitude is dominated by single nucleon production.  
Instead, even subdominant multinucleon contributions have to be under control. 
The net result is that the nuclear form factor is multiplied by the factor $1+a'\frac{q^2}{m_\pi^2}$ and the effects of an uncertainty in the value of $a'$ can not be distinguished from slight changes in the value of $R_A$. 
Instead of measuring $R_A$ one measures  $R^{\rm eff}$, related to the correct value $R_A$ by
$$
	{\left(R_A^{\rm eff}\right)}^2 = R_A^2+6\frac{a'}{m_\pi^2}.
$$
Taking the square root of this equation we get $R_A^{\rm eff} = R_A\left(1+6\frac{a'}{m_\pi^2}\right)^{1/2} \approx R_A-3\frac{a'}{m_\pi^2R_A}$.
The second term hence corresponds to the uncertainty in $R_A$ introduced by the sources discussed above. 
Assuming the error claimed by Ref.~\cite{CBMAMI}, \IE, 0.03 fm, the corresponding value for $a'$ is $a' = (0.03\rm\ fm)\frac{m_\pi^2R_A}{3} \approx 0.03$.
Thus, one needs to know and include  all higher order corrections at the 3\% level in order to accomplish the claimed error in the neutron skin. 
The neglected effects, \EG, the pion charge exchange, are 30-40\% corrections for light nuclei~\cite{EW}, and the effects of using different inner optical potential wave functions on angular distributions can be very large~\cite{Miller:1974nm,Keister:1978ey}. 
Thus the theory used to extract the neutron radius does not have the required accuracy.

The error given in Ref.~\cite{CBMAMI} is optimistic, because it ignores effects that are important for a precision extraction of the neutron radius.
A detailed calculation of these amplitudes is in progress.

\begin{acknowledgments}
A. G. thanks the Indiana University Nuclear Theory Center for its hospitality. 
The work of C. J. H. was supported  by the U. S. Department of Energy Office of Science, Office of Basic Energy Sciences program under Award Numbers DE-FG02-87ER40365 (IU) and DE-SC0008808 (NUCLEI SciDAC Collaboration).   
The work of G. A. M. was supported by the U. S. Department of Energy Office of Science, Office of Basic Energy Sciences program under Award Number DE-FG02-97ER-41014. 
\end{acknowledgments}

\bibliographystyle{apsrev}

\end{document}